\documentclass[prl,twocolumn,showpacs,amsmath,amssymb]{revtex4}
\usepackage{bm}
\usepackage{graphicx}
\begin{document}
\title{Relating supercooling and glass-like arrest of kinetics for phase separated systems: studies on doped CeFe$_2$ and (La,Pr,Ca)MnO$_3$.}
\author{Kranti Kumar, A. K. Pramanik, A. Banerjee and P. Chaddah} 
\affiliation{UGC-DAE Consortium for Scientific Research(CSR)\\ University Campus, Khandwa Road. Indore 452017. India.}
\author{S. B. Roy}
\affiliation{Magnetic and Superconducting Materials Section, Raja Ramanna Centre for Advanced Technology, Indore 452013, India.} 
\author{S. Park, C. L. Zhang, and S-W. Cheong}
\affiliation{Department of Physics \& Astronomy, Rutgers University, Piscataway, New Jersey 08854, USA}\date{\today}
\begin{abstract}
Coexisting ferromagnetic and antiferromagnetic phases over a range of temperature as well as magnetic field have been reported in many materials of current interest, showing disorder-broadened 1st order transitions. Anomalous history effects observed in magnetization and resistivity are being explained invoking the concepts of kinetic arrest akin to glass transitions. From magnetization measurements traversing novel paths in field-temperature space, we obtain the intriguing result that the regions of the sample which can be supercooled to lower temperatures undergo kinetic-arrest at higher temperatures, and vice versa. Our results are for two diverse systems viz. the inter-metallic doped CeFe$_2$ which has an antiferromagnetic ground state, and the oxide La-Pr-Ca-Mn-O which has a ferromagnetic ground state, indicating the possible universality of this effect of disorder on the widely encountered phenomenon of glass-like arrest of kinetics.     
\end{abstract}

\maketitle
Magnetic field (H) and temperature (T) induced broad first order phase transitions (FOPT), and the resulting spatial phase separation is being considered to be responsible for the functional properties of various materials of current interest like colossal magnetoresistance (CMR) materials \cite{dag}, giant magnetocaloric materials, magnetic shape memory alloys etc. Most of these are multi-component systems, having quenched disorder.  In an early theoretical work, Imry and Wortis \cite{imry} showed that such quenched disorder would give rise to broad transition with spatial distribution (or landscape) of the phase transition line (H${_C}$, T${_C}$) across the sample. The first visual realization of such a landscape was provided by Soibel et al. \cite{soibel} for the vortex lattice melting transition. A similar visual realization for an antiferromagnetic (AFM) to ferromagnetic (FM) transition, in doped CeFe$_2$, was provided by Roy et al. \cite{roy} for the FOPT being caused by variation of either T or H. Similar phase separations observed in many other systems of current interest arise from the landscape of free-energy densities and a spread of local (H${_C}$, T${_C}$) values across the sample. The large number of (H${_C}$, T${_C}$) lines would thus form a band. The spinodal lines corresponding to the limit of supercooling (H*, T*) and corresponding to the limit of superheating (H**, T**) would also be broadened into bands for samples with quenched disorder \cite{manekar}. Each of these bands corresponds to a quasi-continuum of lines; each line represents a region of the disordered sample, and it is imperative to understand the nature of these bands to explain the static as well as dynamic properties of the phase separated systems.    

In a parallel development, the kinetics of phase transformation  and its effect on the observed static properties in the disorder-broadened FOPTs are being seriously investigated \cite{manekar,singh,chatto1,chatto2,sharma,ghive}. It is believed that in various materials the kinetics is actually arrested (on experimental or laboratory time scales) and a `glass' is formed \cite{ref1}; demonstrating apparently anomalous variation in physical properties as the H-T plane is spanned. In addition to CeFe${_2}$, these effects have been seen in Gd${_5}$Ge${_4}$ \cite{chatto1}, La-Pr-Ca-Mn-O (LPCMO)\cite{sharma,ghive}, Nd-Sr-Mn-O \cite{kuwa,tokura1}, Nd$_7$Rh$_3$ \cite{sampath} (see also references [5, 6]). The FOPT can be fully or partially arrested at low temperatures; the arrest occurs as one cools, and this frozen state gets `de-arrested' over a range of temperature (which depends on the H) as one warms \cite{chatto2,sharma,ghive}. If such a kinetic arrest were to occur below a (H${_K}$, T${_K}$) line in the pure system (analogous to glass transition line (T$_g$) for topological glasses), the disordered system would have a (H${_K}$, T${_K}$) band formed out of the quasi-continuum of (H${_K}$, T${_K}$) lines. Each line in this (H${_K}$, T${_K}$) band representing a local region of the sample would have its conjugate in the (H*, T*) band.

In this letter, we seek a correlation between the ordering of these quasi-continuum of lines in (H*, T*) and (H${_K}$, T${_K}$) bands from magnetization (M) measurements, and show that the type of correlation has profound effect on the behavior of M for two divergent systems. The most intriguing outcome of this study is the unified observation of an \emph{anti-correlation} implying that the region having higher T* has lower T$_K$ for a pollycrystalline sample of disordered C15 Laves phase compound Ce(Fe, 2\%Os)$_2$ having AFM ground state as well as for a single crystal of charge ordered, CMR manganite, LPCMO having FM ground state.

We first take a case where the high T state is FM and the low T ground state is AFM. Since in a higher field FM will exist over a larger T-region, the (H*, T*)  band falls to lower T with rising H. 
\begin{figure}[hb]
	\centering
		\includegraphics{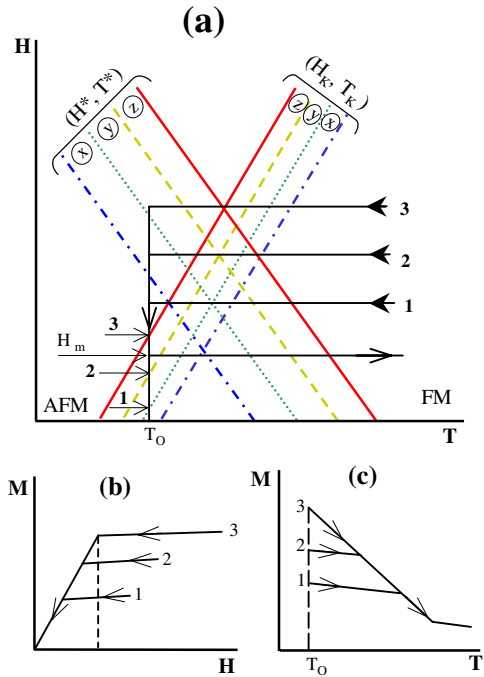}
	\caption{(color online) Schematics of H-T phase diagrams as well as the M-H and M-T curves after cooling in different fields for the case with AFM ground state. (a) H-T diagram with the anti-correlated bands showing only 3 regions (viz. x, y, z) out of the N regions represented by the quasi-continuous (H*, T*) band. The corresponding 3-regions and the dividing lines in the (H$_K$, T$_K$) band appear in reverse order because of anti-correlation. For the first measurement protocol, after FC in different H to T$_O$ following paths 1-3 the H is reduced to zero. The variation in M resulting from this isothermal decrease in H is sketched in (b). Steep decrease in M starting at progressively higher H is due to de-arrest for the respective FC paths starting at higher H, as depicted by horizontal arrows in (a). (c) depicts the next measurement protocol, after FC along paths 1-3 of (a) the H is isothermally lowered to that of H$_m$, the M is measured while increasing the T from T$_O$. The abrupt change in the slope of M occurring at progressively higher T for the FC paths of lower H indicates the increase in the T at which de-arrest begins.}
	\label{fig:Fig1}
\end{figure}
We take the (H${_K}$, T${_K}$) band to be below the (H*,T*) band at zero field (to ensure that zero-field cooled (ZFC) state is fully transformed AFM), and to rise to higher H as T is raised, and cross the (H*, T*) band consistent with the earlier work on Al-doped CeFe$_2$. This is shown in the schematic Figs. 1(a) \& 2(a) where only 4 out of the quasi-continuum of N-lines are depicted to exemplify 3 regions of the sample (viz. x, y and z).
When the high-T end of the (H*, T*) band and the low-T end of the (H${_K}$, T${_K}$) band correspond to the same local region, the two bands are said to be anti-correlated as indicated in the Fig. 1(a); whereas the reverse is the case for the correlated bands as shown in Fig. 2(a). For our study, we prepare the low temperature state of the sample by cooling it from high T to T$_O$ in different H shown by paths 1-3 in Fig. 1(a) and 2(a).
\begin{figure}[ht]
	\centering
		\includegraphics{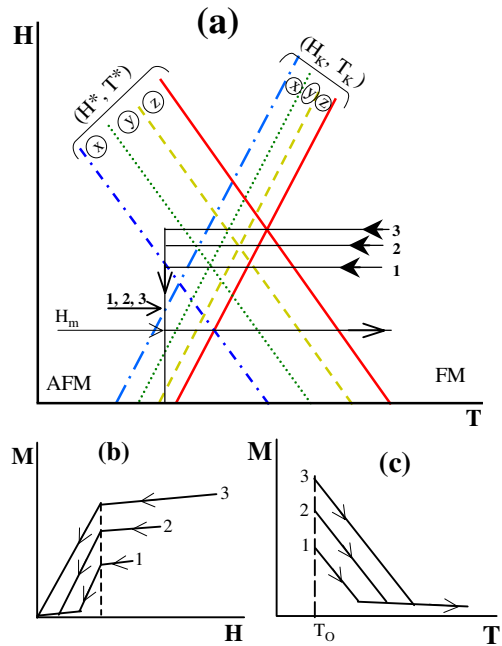}
	\caption{(color online) (a), (b) and (c) are the counter parts of figures 1(a), 1(b) and 1(c) respectively, for the case of AFM ground state when the bands are correlated. It is very clear that M-H and M-T behavior for the correlated bands shown in (b) and (c) are very distinct from that of the anti-correlated bands shown in Figs. 1(b) and 1(c).}
	\label{fig:Fig2}
\end{figure}
Then we simulate the variation of M for two different measurement protocols. In the first case, we measure M  while reducing the field to zero and in the second case we reduce the field to a non-zero low value H$_m$ with T held at T$_O$, and then measure M while increasing T.

We now study the variation of M for different FC states assuming anti-correlated bands as depicted in Fig. 1(a). Cooling along path-3, the entire sample gets kinetically arrested in the high-T FM state as T$_O$ is approached because for this path, (H${_K}$, T${_K}$) lines of all the regions are at a higher T than the corresponding (H*, T*) lines. Cooling in lower H along path 2, the 1st order transformation of the regions in group z to AFM state will start before they can get arrested since the (H${_K}$, T${_K}$) lines between the dashed and the continuous lines of the associated regions are at lower T than the corresponding lines of the (H*, T*) band. However, the regions in group y and x are arrested in the FM phase. Similarly, the regions in group z and y are transformed when the sample is cooled along path-1 because they encounter the associated (H*, T*) lines while cooling before the corresponding kinetic arrest lines in the (H${_K}$, T${_K}$) band, however, the regions in group x will be in arrested FM phase. Thus field cooling (FC) in progressively higher fields renders more frozen FM phase. As we cool in higher H, the additional region arrested move from y to z. Now for the first measurement protocol, the field is isothermally reduced to zero at T$_O$ for all the FC paths. The arrested FM phase will start getting de-arrested to the equilibrium AFM phase at the fields shown by the horizontal arrows for the respective paths during the field reduction at T$_O$. For path-3, the de-arrest will start as soon as the (H${_K}$, T${_K}$) band is entered and successively the regions in group z, y and x will be transformed as the (H${_K}$, T${_K}$) band is traversed in the opposite sense during the reduction of field to zero. However, during the reduction of H for paths 2, the de-arrest will start when the dashed line of (H${_K}$, T${_K}$) band is encountered since the region z has already transformed while cooling, only regions in group y and x were arrested. Similarly for path-1 only the regions in x is arrested and their transformation will start at the dotted line as zero H is approached. The variation of M for this isothermal decrease in H is schematically shown in Fig. 1(b) where the sharp decrease in M at progressively higher H for the respective FC paths resulted from the starting of de-arrest at successively higher H as shown in Fig. 1(a). For the second measurement protocol, as the H is reduced, part of the regions of group z will get de-arrested by the time the measurement field (H$_m$) is approached for the FC state of path-3. With the increase in T the de-arrest will continue and the M will decrease monotonically as schematically shown in Fig. 1(c). Whereas such monotonic decrease will start only at a higher T for the FC state of path-2 and the M will merge with the path-3 when the de-arrest of the regions in group y start after encountering the dashed line during heating. Similarly, the de-arrest for the FC path-1 will start at still higher temperature when the dotted line is encountered and the M will merge with the rest as shown in Fig. 1(c).

On the contrary, for correlated bands shown in Fig. 2(a), all the FC paths will have regions of group x in arrested phase at T$_O$, additionally the FC path-2 will have regions of group y and the path-3 will have regions of group y and z in arrested phase. Hence the de-arrest will simultaneously start for all the paths as soon as the dash-dotted line is encountered when the H is decreased at T$_O$ shown by the horizontal arrow. Thus for the M vs. H at T$_O$ the decrease in M will start at the same field for all the FC paths. However all the paths having different amount of the arrested phase, the de-arrest will continue down to different value of H and M will independently go to zero when H is reduced to zero as schematically shown in Fig. 2(b). It may be noted that this in complete contrast with the Fig. 1(b). Similarly, when the temperature is increased after reducing the H to H$_m$, the de-arrest for all will start at T$_O$. However, the decrease in M with higher slope will continue up to successively higher T for higher FC path because of larger fraction of arrested FM phase as shown in Fig. 2(c). This M-T behavior is also significantly different from the case of anti-correlated band shown in Fig. 1(c). 

The second case is for systems having AFM high-T state and low-T FM ground state. A detailed discussion can be found in reference \cite{chaddah}. For this case, the low-T state is prepared by cooling in different fields including zero-field to T$_O$. The M is measured in two protocols: (i) by isothermally increasing H at T$_O$, and (ii) by raising the field value to H$_4$ and then increasing the T. The schematic of the resulting M-H and M-T are reproduced in Figs. 4(a) \& 4(c) respectively for only the anti-correlated bands. It can be easily visualized that M-H or M-T behaviors are in complete contrast for correlated bands, analogous to the previous case \cite{chaddah}.

To present experimental test for both the above mentioned cases magnetization measurements are carried out on 2\% Os-doped CeFe$_2$ and LPCMO. The details about sample preparation and charecterization can be found in reference \cite{roy2} for polycrystalline Ce(Fe, 2\%Os)$_2$ and in reference \cite{sharma} for LPCMO.  The M measurement is performed on a 54.5 mg Ce(Fe, 2\%Os)$_2$ sample using a commercial 14 Tesla Vibrating Sample Magnetometer (Quantum Design, PPMS-VSM) and the M measurements for single crystal of LPCMO is performed with a SQUID magnetometer (Quantum Design, MPMS)  

Fig. 3(a) shows M-H curves for Ce(Fe, 2\%Os)$_2$ obtained after FC in various fields to 2K, 4K and 20K. \begin{figure}[ht]
	\centering
		\includegraphics{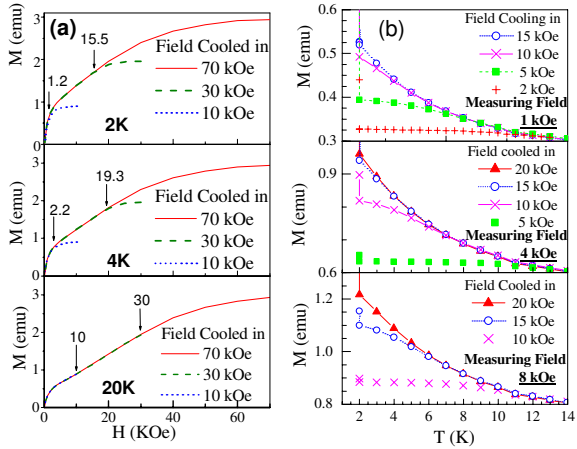}
	\caption{(color online) (a) Shows the M-H of Ce(Fe, 2\%Os)$_2$ after cooling in 70, 30 and 10 kOe to 2K, 4K and 20K then reducing the field to zero. The vertical arrows show the merger of 70 kOe curve with that of 30 kOe and of 10 kOe curve with the rest and gives the H values for the respective merger. (b) shows the M-T after FC in different H to 2K where H is isothermally reduced to the measuring fields (1 kOe, 4 kOe and 8 kOe).}
	\label{fig:Fig3}
\end{figure}
It can be clearly seen that field decreasing curves for 2K and 4K are similar to schematic of Fig. 1(b). The merger behavior remains qualitatively similar for both 2K and 4K, though the merger fields increase for 4K which is obvious from the schematic of Fig. 1(a). Since there is no arrested phase at 20K the M-H curves overlap. Fig. 3(b) shows the M-T behavior for Ce(Fe, 2\%Os)$_2$ after FC to 2K in different H and then after reducing the H to various measurement fields, M is measured as the T is raised from 2K. It can be noticed that the T of the merger of M of a higher-H FC state with the next FC state increases as the cooling field decreases since the de-arrest for a lower FC state start at a higher T which is consistent with Fig. 1(a). Hence, both the M-H and M-T behavior for Ce(Fe, 2\%Os)$_2$ is qualitatively similar to Figs. 1(b) and 1(c) respectively indicating that the bands are anti-correlated for this case. 

Fig. 4(b) shows the M-H for LPCMO where the sample is cooled in different fields including zero to 5K and then the H is raised. It is obvious that this behavior of M is qualitatively similar to Fig. 4(a) where the de-arrest of the higher-H FC paths start at higher H and merges with the ZFC curve indicative of the anti-correlated bands. For M-T measurement, the LPCMO is cooled to 5K in different H, then H is increase to 10 kOe and the M is measured while increasing T. It can be observed from the resulting M-T curves (Fig. 4(d)) that the de-arrest for higher-H FC paths stars at higher-T and merges with the ZFC curve similar to the schematic of Fig. 4(c) confirming anti-correlated bands for this system as well (see also discussions in reference \cite{chaddah}).
\begin{figure}[ht]
	\centering
		\includegraphics{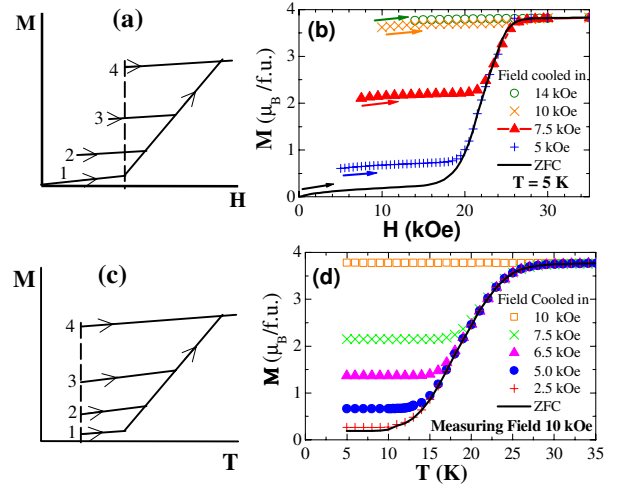}
	\caption{(color online) The M-H and M-T behavior for the case with FM ground state. (a) Schematic M-H after FC in different H for anti-correlated bands. (b) shows the corresponding measurement on LPCMO after FC in different H to 5K and then increasing the H. (c) schematic M-T  for anti-correlated bands after FC in different H, then raising H to H$_4$ and measuring M as T is raised. (d) shows the measured M-T on LPCMO after FC in different H including ZFC to 5K, raising H to 10 kOe at 5K and then increasing the T.}
	\label{fig:Fig4}
\end{figure}

From these experiments we can conclude that both the disordered C15 Laves phase compound Ce(Fe, 2\%Os)$_2$ and disordered CMR manganite LPCMO unambiguously show anti-correlation between the kinetic arrest band and the supercooling band.  \emph{Is this anti-correlation universal and can its microscopic origin be understood? Does it have implications for the general problem of glass formation?} Our results presented here are for two very diverse systems whose commonality is the coexistence of FM and AFM regions, otherwise separated by a 1st order transition, over a broad range of T and H. Doped CeFe$_2$ is an intermetallic with both FM and AFM phases showing metallic resistivity. The low-T ground state is AFM, and the sample used is polycrystalline. La-Pr-Ca-Mn-O is an oxide with the FM phase showing metallic resistivity, while the AFM phase is insulating. The low-T ground state is FM, and the sample used is a single crystal. This could be indicating the possible universality of this effect of disorder, and similar features could be studied in many diverse systems like Gd$_5$Ge$_4$, Nd-Sr-Mn-O, Nd$_7$Rh$_3$, etc. While the kinetic arrest associated with the glass transition is usually studied as a function of T alone, we have addressed systems where two control variables viz. T and H are being used to cause arrest (or de-arrest). Varying H is experimentally straightforward, and studies in these systems could allow a new insight into the widely encountered phenomenon of glass-like arrest of kinetics. Assuming that the results found in this study have a broader validity, we venture to suggest that the glass formation becomes easier in complicated molecules, indicating that kinetics is more easily hindered and arrested as near-neighbor disorder rises. On the other hand, such disorder would lower the phase transition temperature. At a naive level, this could explain the above inference. It needs to be mentioned here that we have not included relaxation effects in our present study which will be dealt in a detailed paper.

	DST, Government of India is acknowledged for funding the VSM used for the measurement. AKP acknowledges CSIR, India for fellowship. Work at Rutgers was supported by NSF-DMR-0405682.       

{ }


\begin{thebibliography}{ }
\bibitem[11]{ref1}The metastable kinetically arrested (or glass) state is different from the metastable supercooled state in that the latter (but not the former) will undergo a metastable to stable transformation on lowering of temperature. In other words, the relaxation time decreases with the decrease in T for a supercooled state whereas the relaxation time increases with the decrease in T for the glassy state.
\end{thebibliography}
\end{document}